\documentclass[preprint,
preprintnumbers,amsmath,amssymb,nofootinbib]{revtex4}

\usepackage{etex}
\usepackage{graphicx}
\usepackage{dcolumn}
\usepackage{bm}
\usepackage{amssymb,amsthm,amscd,amsbsy,array}\usepackage{amsmath,dsfont}

\usepackage{graphics,xcolor}

\usepackage{color}

\newcommand{\UU}{\mathrm{U}}
\newcommand{\half}{{\scriptstyle{\frac{1}{2}}}}
\def\2{{\half}}

\def\parag{\hfil\break} 
\def\kikezd{\parag\underbar}

\def\p{{\partial}}

\def\beq{\begin{equation}}
\def\eeq{\end{equation}}
\def\beqa{\begin{eqnarray}}
\def\eeqa{\end{eqnarray}}

\def\barray{\left(\begin{array}}
\def\earray{\end{array}\right)}
\def\barraynb{\begin{array}}
\def\earraynb{\end{array}}
\def\IR{{\mathds{R}}} 
\def\IF{{\mathds{F}}}
\def\IZ{{\mathds{Z}}}

\def\smallover#1/#2{\hbox{$\textstyle\frac{#1}{#2}$}} %

\def\tgamma{{\widetilde{\gamma}}}

\newcommand{\cF}{\mathcal{F}}

\def\smallcirc{{\raise 0.5pt \hbox{$\scriptstyle\circ$}}}
\def\aand{{\quad\text{\small and}\quad}}

\def\where{{\quad\text{\small where}\quad}}

\def\ie{{\;\text{\small i.e.}\;}}
\def\ie,{{\;\text{\small i.e.,}\;}}

\def\benu{\begin{enumerate}}
\def\eenu{\end{enumerate}}
\def\bitem{\begin{itemize}}
\def\eitem{\end{itemize}}

\def\besub{\begin{subequations}}
\def\esub{\end{subequations}}

\def\?{{\,\gb{\fbox{\texttt{??}}\;}}\,}

\def\Rarrow{{\quad\Rightarrow\quad}}

\usepackage[colorlinks=true, pdfstartview=FitV, linkcolor=blue, citecolor=blue, urlcolor=blue]{hyperref} 

\newcommand{\gb}{\quad\colorbox{green}}

\newenvironment{redtext}{\color{red}}
{\ignorespacesafterend}
\newenvironment{bluetext}{\color{blue}}{\ignorespacesafterend}
\newenvironment{greentext}{\color{green}}{\ignorespacesafterend}
\newenvironment{magentatext}{\color{magenta}}{\ignorespacesafterend}
\newenvironment{cyantext}{\color{cyan}}{\ignorespacesafterend}
\newenvironment{orangetext}{\color{orange}}
{\ignorespacesafterend}

\newcommand{\bmagenta}{\begin{magentatext}}
\newcommand{\emagenta}{\end{magentatext}}
\newcommand{\bcyan}{\begin{cyantext}}
\newcommand{\ecyan}{\end{cyantext}}
\newcommand{\bblue}{\begin{bluetext}}
\newcommand{\eblue}{\end{bluetext}}
\newcommand{\bred}{\begin{redtext}}
\newcommand{\ered}{\end{redtext}}

\newcommand{\bgreen}{\begin{greentext}}
\newcommand{\egreen}{\end{greentext}}
\newcommand{\borange}{\begin{orangetext}}
\newcommand{\eorange}{\end{orangetext}}

\numberwithin{equation}{section}
\renewcommand{\theequation}{\thesection.\arabic{equation}}
\let\ssection=\section
\renewcommand{\section}{\setcounter{equation}{0}\ssection}

\newcommand{\bigbox}[1]{\fbox{%
\rule[-20pt]{0pt}{45pt}$\;\;\displaystyle{#1}\;\;$}
}

\newtheorem{thm}{Theorem}
{Corollary}
\newtheorem{defin}{Definition}
\newtheorem{prop}{Proposition}
\newtheorem{lemma}
{Lemma}[section]

\newcommand{\AB}{Aharonov-Bohm\;}

\begin{document}


\title{Prequantisation from the path integral viewpoint\footnote{Slightly reworded, rearranged and extended version of one originally published in the Proceedings of the {\sl Int. Conf. on Diff. Geom. Meths. in Math. Phys.}, Clausthal'1980, Doebner (ed). Springer Lecture Notes in Math. \textbf{905}, p.197-206 (1982). Marseille preprint CPT-80-P-1230.}
}

\author{
P. A. Horv\'athy$^{1,2,3}$\footnote{mailto:horvathy@univ-tours.fr }
}

\affiliation{
${}^1$ CNRS; Centre de Physique Th\'eorique, F-13288 Marseille, Cedex 2 (France)
\\
${}^2$ Ist. Fis. Mat. dell'Universit\`a, Via Carlo Alberto 10, I-10123 Torino (Italy)
\\
${}^3$ Current address:  Institut Denis-Poisson CNRS/UMR 7013 - Universit\'e de Tours - Universit\'e d'Orl\'eans Parc de Grammont, 37200, Tours, (France)
\\
}
\date{\today}
\begin{abstract}
The quantum mechanically admissible definitions of the factor $\exp\big[(i/\hbar)S(\gamma)\big]$  in the Feynman integral are put in bijection with the prequantisations of Kostant and Souriau. The different allowed expressions of this factor -- the inequivalent prequantisations -- are classified. The theory is illustrated by the Aharonov-Bohm experiment and by identical particles.
\end{abstract}

\maketitle

\tableofcontents

\section{Introduction}

In \cite{HPAix79} an attempt was made to use the {geometric techniques} of  Kostant and Souriau
\cite{Kostant,SSD} (``K-S theory") to study {path integrals}. The method was then applied to a Dirac monopole and to the \AB experiment. In this note we generalize those results. We show indeed that a general symplectic system is \emph{quantum mechanically admissible} (Q.M.A.) iff it is \emph{prequantisable} in the Kostant-Souriau sence \cite{Kostant,SSD,SimmsWoodhouse} with transition functions which depend only on space-time variables. 

If the configuration space is not simply connected, then a given classical system may admit different, inequivalent quantisations~\footnote{Early references include the pioneering work of Souriau who used his ``quantification g\'eom\'etrique''  framework \cite{JMS67, SSD}. Inequivalent quantizations  were discussed later for path integrals in  \cite{Schulman,Laidlaw, Dowker72,SchulmanPathInt}.
  Attempts to combine the two approaches are found in  \cite{SimmsAix79, HPAix79,HPAPLA,HPClausthal,HoMoSu}. },
implied by inequivalent prequantisations. 
The main result is the content of in Theorem \ref{concompchar} announced  in sec.\ref{Classif}, obtained by spelling out the classification scheme implicitly recognised by Kostant \cite{Kostant}, and by Dowker \cite{DowkerAustin}, is illustrated in FIG.\ref{classiffig}. Its proof  \cite{ackKollar} is presented in the Appendix. 

\begin{figure}
\includegraphics[scale=.4]{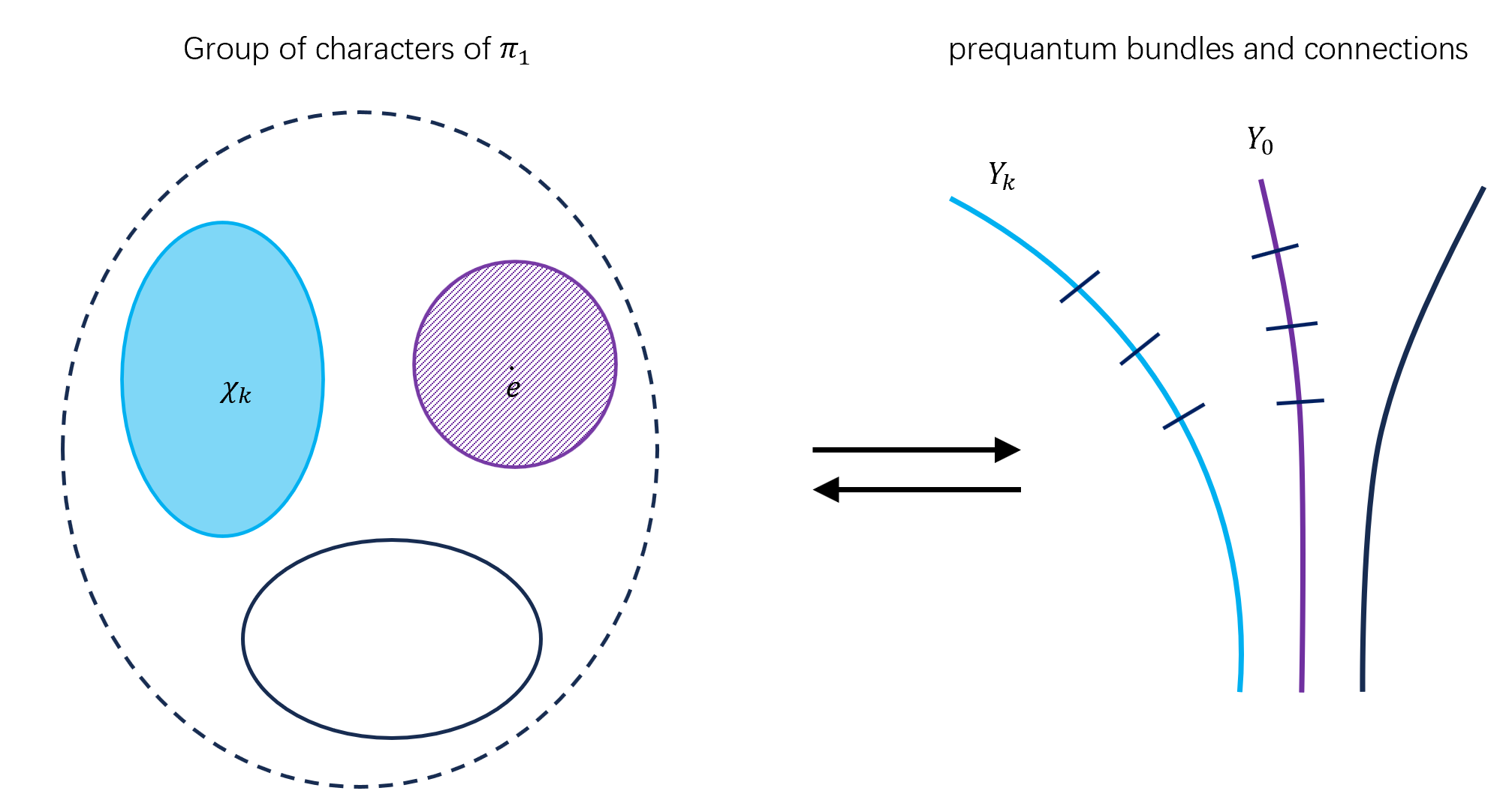}
\caption{\textit{\small 
In $d\geq3$ spatial dimensions, the inequivalent quantisations 
 correspond to the characters of the fundamental group of the configuration space, $\pi_1(Q)$, \eqref{chiofg}. The topologically distinct bundles correspond to the components of the character group and can be labelled by choosing (non canonically) a character $\chi_i$  in its respective component. The inequivalent connections on a chosen bundle are in bijection with the characters in the identity component
 which correspond physically to curl-less ``\AB- type'' vector potentials.
}
\label{classiffig}
}
\end{figure}

The fundamental concept  is the ``Feynman factor''\footnote{The idea first was put forward by  Dirac \cite{Dirac33} and then further developed by Feynman in his PhD thesis \cite{Feynman42}.  
On path  integrals see, e.g., \cite{FeynmanHibbs,SchulmanPathInt}.}
\beq
\cF(\gamma) = 
\exp\left[\smallover{i}/{\hbar} S(\gamma)\right]\,,
\label{Ffactor}
\eeq
where $S(\gamma)$ is the classical action calculated along  the path $\gamma$.

The aim of this note is to provide physicists with an introduction to (geometric) {prequantisation} and to 
contribute to its physical interpretation \cite{SimmsWoodhouse,SimmsAix79}.
 
\goodbreak

\section{Quantum mechanically Admissible (Q.M.A.) Systems}\label{QMAS}

Let us restrict ourselves to classical system $(E,\sigma)$
with evolution space $E = T^{*}Q\times \IR$, where $Q$ is the configuration space (where we follow Souriau's framework \cite{SSD}). $E$ is endowed with a presymplectic structure of the form
\beq
\sigma = d\theta_0 + e\IF\,,
\label{sigmaform}
\eeq
where $\theta_0$, obtained by  restricting the canoncial $1$-form of 
$T^{*}(Q\times \IR)$  to the energy surface $H=H_0(q,p,t)$ , describes a free system \cite{SimmsWoodhouse,SimmsAix79}.
  $\IF$, a closed $2$-form on space-time $X =Q\times \IR$, represents the external electromagnetic field to which our particle is coupled by the constant $\underline{e}$ \cite{SSD}.
  
If the system admits a Lagrangian, then $\sigma = d\theta$ 
where $\theta$ is the  Cartan form of the variational system \cite{SSD,HPAix79}. The Hamiltonian action is obtained by  integrating alongs paths  $\gamma$ in phase (more precisely in evolution) space, whose initial and final points project to the same $x=(q,t)$ and $x'=(q',t')$ in space-time $X$ in the Feynman factor \eqref{Ffactor},
\beq
S(\gamma)=\int_\gamma\theta\,.
\label{claction}
\eeq

\begin{defin} $(E,\sigma)$ is \emph{quantum mechanically admissible} (Q.M.A.) system iff $E$ can be covered by  a collection $\left\{U_j,\, \theta_j\right\}$ of pairs of contractible open subsets $U_j$ and $1$-forms $\theta_j$ defined on them such that  for any $\gamma \subset U_j\cap U_k$ 
 we have,
\beq
\exp\left[\smallover{i}/{\hbar}\int_{\gamma}\theta_j\right] 
= C_{jk}(x,x')\exp\left[\smallover{i}/{\hbar}\int_{\gamma}\theta_k\right]\,,
\label{ijintegral}
\eeq
where the unitary complex factors $C_{jk}$ depend only on the initial and end points $x,\,x'$ of $\gamma$ but not on $\gamma$ itself.
\label{QMA}
\end{defin}

When \eqref{ijintegral} is satisfied, then the Feynman factors \eqref{Ffactor} which correspond to $\theta_j$ resp $\theta_k$  are related by unobservable phase factors.
The idea comes from recalling that adding a total derivative to the Lagrangian (which amounts to adding an exact $1$-form to the Cartan form $\theta$) changes both $\cF$ and the wave function by an unobservable phase factor \cite{HPAix79}.

\begin{figure}
\includegraphics[scale=.6]{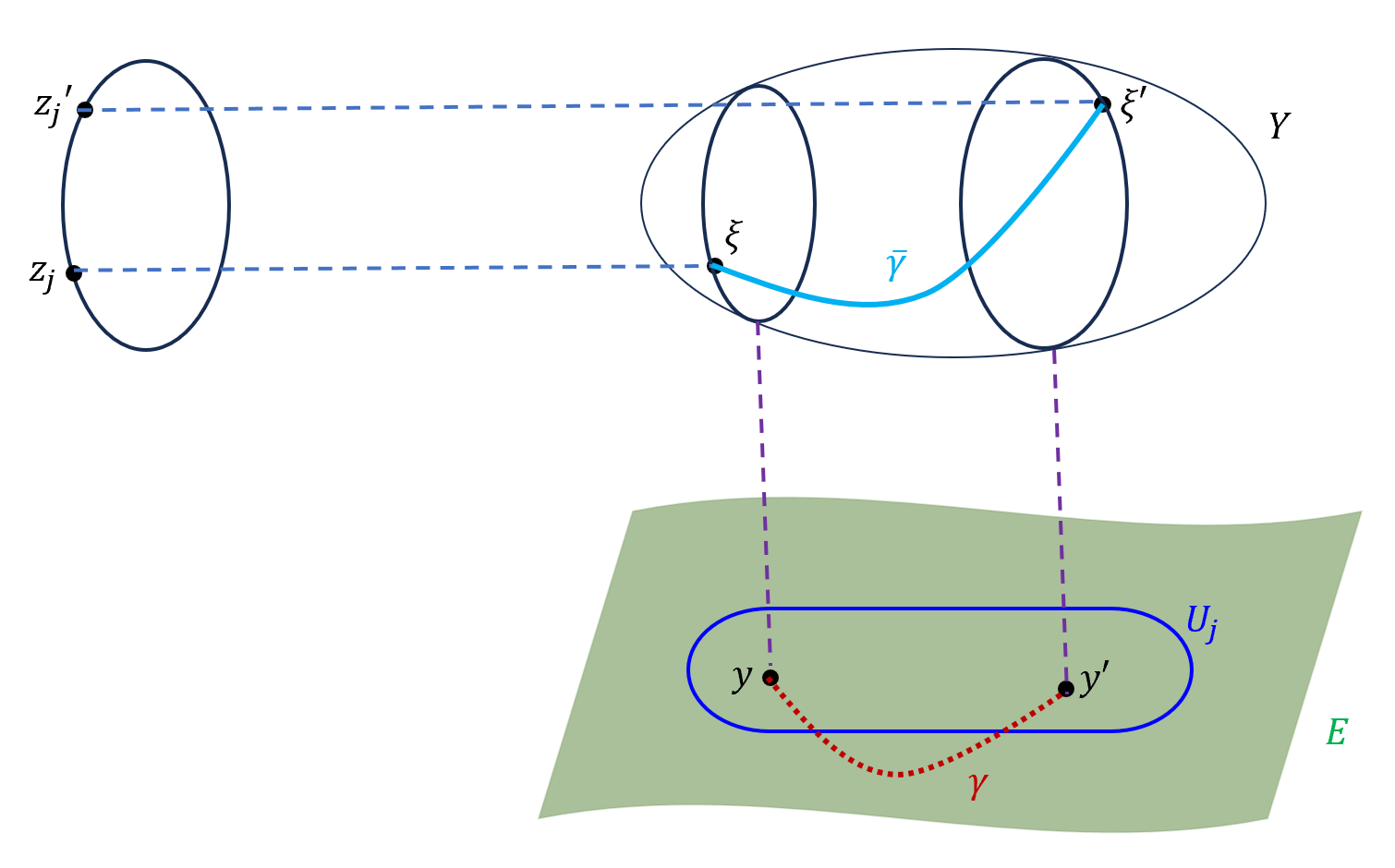} 
\caption{\textit{\small The Feynman factor \eqref{Ffactor} for a path $\gamma$ whose initial and end points $y,\,y'$ (but necesarilly the entire $\gamma$)
 belong to a chosen contractible subset $U_j$, 
  is \underline{defined} by \underline{lifting $\gamma$ horizontally} to the prequantum bundle $Y$ as $\bar{\gamma}$ and then reading off from the change of the vertical coordinate in a local trivialisation} \cite{HPAix79,SimmsAix79}.
\label{Fffig}
}
\end{figure}

By the Weil lemma the consistency condition \eqref{ijintegral} is equivalent to requiring that the system be prequantisable \cite{Kostant,SSD,SimmsWoodhouse}, 
\beq
\frac{1}{2\pi\hbar}\int_S\sigma \in \IZ
\label{prequantcond}
\eeq
for any $2$-cycle $S$ in space-time \footnote{For a Dirac monopole this condition implies the quantization of the electric charge \cite{HPAix79}, and for a spinning particle it implies the analogous quantization of the spin \cite{SSD}. Similar arguments apply to the classical isospin \cite{Torino81}.}.

The bundle framework allows us to define the Feynman factor \eqref{Ffactor} is \emph{any} curve $\gamma$.
To this end, we proceed as follows. The transition functions of the $\UU(1)$ bundle $Z_{jk}: U_{j} \cap U_k \to U(1)$ satisfy, with the locally defined Cartan forms, 
\beq
\theta_j - \theta_k = \frac{dZ_{jk}}{iZ_{jk}}
\Rarrow
C_{jk}(x,x')=\frac{Z_{jk}(x)}{Z_{jk}(x')}\,.
\label{thetajk}
\eeq
Let $\gamma$ be any path in $E$ joining $y
$
and $y' $\, see FIG. \ref{Fffig}. 

 Denote by $\big(Y,\omega,\pi\big)$ a prequantization of $(E,\sigma)$ \cite{Kostant,SSD,HPAix79} where
 \beq
\omega \quad\text{\small locally}\quad
\omega\big|_{U_j}= \theta +\frac{dz}{iz}
\label{prequantform}
\eeq
is the prequantum 1-form \cite{SSD}, defined by $\pi^*\sigma = d\omega$. 
Denote by $\bar{\gamma}$ the \emph{horizontal lift} of $\gamma$ to $Y$,
$\omega(\bar{\gamma})=0$ with $\bar{\gamma}(0) = \xi\in \pi^{-1}(y)$. 
In a local trivialization of the prequantum bundle, $Y \simeq U_j \times \UU(1)$ \cite{SimmsAix79}, 
\beq
\bar{\gamma}(t) = \big(\gamma(t), Z(t)\big)
\where Z(t)=Z_0 \exp \left[-\smallover{i}/{\hbar}\,\int_{0}^t\!\theta_j(\dot{\gamma}(\tau))d\tau \right]\,.
\label{gammalift}
\eeq
Thus the {exponential} of the classical action calculated along the \emph{arbitrary} path $\gamma(t),\, 0\leq t \leq 1$ 
 can be recovered from the horizontal lift as
\beq
\bigbox{
\cF_j(\gamma) =
\exp\left[\smallover{i}/{\hbar}\int_{\gamma}\theta_j\right]=
\frac{Z_j(y)}{Z_j(y')}\,,
}
\label{Ffdef}
\eeq
where $\xi = \bar{\gamma}(y) =\big(y,Z_j(0)\big)$ and
$\xi' = \bar{\gamma}(y') =\big(y',Z_j(1)\big)$.
In conclusion,
\begin{thm} 
$(E,\sigma)$ is Q.M.A. iff it is \emph{prequantisable} with transition functions depending only on the space-time, $X$. 
Then the Feynman factor \eqref{Ffactor} is defined for \underline{any curve $\gamma$} whose end points 
 lie  in a single  $U_j$ as shown in FIG.\ref{Fffig}. 
 The consistency relation \eqref{ijintegral} is satisfied. 
\label{factorglueing}
\end{thm}

Our approach, illustrated by FIG.\ref{Fffig},
is then analogous to that followed for monopoles, where the particle Lagrangian is singular along a ``Dirac string'' --- which however can be cured  by lifting the problem to a $\UU(1)$ bundle \cite{HPAix79,WuYangLag,BalaMarmo}.
This allows us  to {define} the factor $\cF(\gamma)$ also for paths which do \emph{not} lie entirely in a fixed  contractible subset $U_j$ by glueing together local  expressions.
 Intuitively, the clue is that ``the exponential of the action may behave better as the action itself''.

\section{Geometric expression for the integrand}\label{GeoInt}

Now we present a coordinate-free form to the integrand in Feynman's expression. Following a suggestion of Friedman and Sorkin \cite{FriedmanSorkin}, let us consider an \emph{arbitrary} (not necessarily horizontal) path $\tgamma$ in $Y$ which projects onto $\gamma$. Write
$\tgamma(0) = \xi \simeq (y,Z_j),\,\tgamma(1) = \xi' \simeq (y',Z_j')$ for the end points, and let $\omega$ be the prequantum form \eqref{prequantform}.
\goodbreak

\begin{lemma}
\beq
\exp\left[\frac{i}{\hbar} S_j(\gamma)\right]=
\frac{Z_{j}'}{Z_{j}} = \exp\left[\frac{i}{\hbar}\, \int_{\tgamma}\omega\right]\,.
\label{FfYdef}
\eeq
The quotient of the vertical $Z$ coordinates is thus the same for all $\tgamma$  above $\gamma$.
\end{lemma}

Now remember that the wave function can be represented by equivariant complex function on $Y$,  \cite{Kostant,SSD},
\beq
\psi\left(\underline{Z}_Y(\xi)\right)=Z\cdot \psi(\xi)\,,
\label{equivwf}
\eeq
where $\underline{Z}_Y$ denotes the action of $\UU(1)$ on $Y$, rather then mere functions on the configuration space $Q$. The usual wave functions $ \psi_j$ are the \emph{local representants} of these objects, obtained as
\beq
\psi(\xi) = Z_j\cdot \psi_j(x),
\where \xi \in \pi^{-1}(U_j) \,.
\label{locwf}
\eeq
Thus we get finally the geometric expression for the time evolution,
\beq
\bigbox{
\left(U_{t'-t}\psi\right) (\xi') = \int_Q \displaystyle{\int_{P_{x,x'}}} \!\!{\cal D}\gamma \exp\left[\smallover{i}/{\hbar}\int_{\bar{\gamma}}
\omega\right]\psi(\xi)\,,
}
\label{uppropag}
\eeq
where $\xi=\tgamma(0)$, $P_{x,x'} =\Big\{\gamma \subset E,\,\gamma(0)=(x,\,\cdot\,),\,\gamma(1)=(x',\,\cdot\,)
\Big\}$ denotes the collection of all paths from $x$ to $x'$.
Note that when $\tgamma(1)=\xi'$ is held fixed, then
$$
\exp\left[\smallover{i}/{\hbar}\int_{\bar{\gamma}}
\omega\right]\psi(\xi)
$$
is in fact a function of the projected curve $\gamma$ and is independent of the concrete choice of $\tgamma$ .

\kikezd{Remarks}
\benu
\item
To  define the ``integration measure'' ${\cal D}\gamma$ is a formidable task which would exceed our scope here. An attempt within the geometric quantisation framework was made in \cite{SimmsAix79}.
\item
The introduction of the bundle $(Y,\omega,\pi)$ allows for developping a generalized variational formalism \cite{FriedmanSorkin}
and allows to study conserved classical quantities.
\eenu 

\section{A classification scheme
}\label{Classif}

If the underlying space is not simply connected, then we may have \emph{more than one} prequantisations and thus several inequivalent meanings of the factor $\cF$ in \eqref{Ffactor}. 
The general construction (actually elaborated in  \cite{JMS67}) is found in Souriau book \cite{SSD} . 
Some details are outlined in the Appendix. Here we merely quote his main result :

\begin{thm}
The inequivalent prequantisations are  in 1-1 correspondence with the characters of the first homotopy group.
\label{prequantchar}
\end{thm}

In \cite{HPAix79} we rederived this theorem in the \AB case from  path-integral considerations, noting that we are always allowed to add to the Cartan $1$-form  $\theta$ a closed but not exact $1$-form $\alpha$ \footnote{For physicists, $\alpha=(A_\mu)$ is a curl-less vectorpotential.
} , which, due to non-simply-connectedness, may change the propagator into an inequivalent one. The corresponding character of $\pi_1$ is then
\beq
\chi(g) = \exp\left[\smallover{i}/{\hbar}\oint_{\gamma}\alpha
\right]\,
\label{chialpha}
\eeq
where $\gamma$ is a loop which represents the homotopy class of $g\in\pi_1$.
In the \AB experiment, for example, $\pi_1=\IZ$
and all characters have this form.

This is however not the general situation. A physically interesting counter-example is that of \emph{identical particles} \cite{SSD,DowkerAustin} \footnote{Quantization of a system of identical particles has  been considered in \cite{JMS67,Laidlaw}. Asorey and Boya \cite{AsoreyBoya} studied the case $\pi_1 \simeq \IZ_p$ for $p > 1$.}.

\kikezd{Example}.
Consider two identical particles moving in $3$-space. The appropriate configuration is then \cite{Laidlaw} $Q= \widetilde{Q}/\IZ_2$, where 
\beq
\widetilde{Q} = \IR^3\times\IR^3 \setminus \{q_1 = q_2\}
\label{idpartQ}
\eeq
is the two-particle configuration space with collisions excluded, 
whose homotopy group is $\pi_1=\IZ_2$. The evolution space $E$ is $T^*{Q}\times\IR$ with $\sigma = d\theta_0^{(1)}+d\theta_0^{(2)}$.
$\pi_1=\IZ_2$ has two characters~:
\beq
\chi_1(z)=1
\aand
\chi_2(z)=-1
\eeq
where $z$ is the interchange of two configurations,
$(q_1,q_2) \to (q_2,q_1)$.
Thus we have two prequantum lifts of $\pi_1$ and thus two prequantisations, the first of which is trivial, while the second is \emph{twisted}. The first one corresponds to \emph{bosons}, the second one to \emph{fermions}  
$\chi_2$ is \emph{not} of the form \eqref{chialpha}
\footnote{These statements are valid in at least 3 space dimensions. In the plane, the two-particle homotopy is larger and leads to  \emph{anyons} \cite{Leinaas77,Wilczek82}.}.

The general situation is described by the classification theorem (see the Appendix) says,
 
\begin{thm}
A character of $\pi_1$ can be written as
\beq
\bigbox{
\chi(g) = \exp \left[\frac{i}{\hbar}\,\displaystyle{\sum_{j=1}^{b_1}}\; a_j\!\oint_{\gamma}\alpha_j\right]\,\chi_k,
}
\label{chiofg}
\eeq
where $g=[\gamma]$ is the homotopy class of the loop $\gamma$,
$a_j\in \IR$ (defined mod $2\pi$). Here $b_1$ is the first Betti number \cite{Betti} which counts the connections on a bundle. The characters $\chi_k$ label in turn the components of the character group.
\label{concompchar}
\end{thm}

At last we get the following refinement of Souriau's theorem in \cite{JMS67,SSD} in \eqref{chiquant}~: 

\begin{thm}
Two prequantum bundles
$Y_{\chi_1}$ and $Y_{\chi_2}$ associated to two 
characters $\chi_1$ and $\chi_2$ are topologically equivalent iff the latter belong to the same component of the group of characters.
Two connection forms on a chosen bundle are labelled by the elements of the connected component of the identity character $\chi\equiv 1$ and correspond, physically, to curl-less ``\AB'' vector potentials. Two such vector potentials define equivalent connections when their fluxes
$\Phi = \oint \!A$ differ by an integer multiple of $2\pi\hbar$. The general situation is illustrated in FIG.\ref{classiffig}.
\label{classifthm}
\end{thm}

For $b_1=1$ and $\alpha_1= \smallover{\Phi}/{2\pi}d\phi$,
 we recover the description of the (Abelian) \AB effect \cite{HPAix79,HoMoSu}. For $\pi_1 =$ Tors = $\IZ_2$ we get instead that of identical-particles in (at least) 3 space dimensions \cite{JMS67,SSD,Laidlaw,DowkerAustin,AsoreyBoya,HoMoSu}.
The Non-Abelian generalization is discussed in \cite{HP-Kollar}.

\vskip-3mm
\begin{acknowledgments} 
 I am indebted to jean-Marie Souriau for hospitality in Marseille. Special thanks are due to J\'anos Koll\'ar for his  help to derive the general classification theorem, outlined in teh Appendix. Discussions are acknowledged also to John Rawnsley. I am grateful also to P-M Zhang for  discussions and his technical help for certain details.
\end{acknowledgments}
\goodbreak


\kikezd{Note added in 2024}. This note was originally published in a conference proceedings  \cite{HPClausthal}. During the long years after its publication I came across several important papers related to the subject but not included in the original, rather succinct note, adressed to specialists in differential geometric methods of physics. In order to put it into wider perspectives and make it more readable to physicists, I decided to make available online. While following as much as I could the original note, I added many commentaries and explanations (in footnotes) as well as a considerably extended reference list [12 - 29]. I have also trasferred to an  Appendix certain geometric details which do not belong to the standard toolbox of physicists.

\bigskip
\appendix
\section{\bf Classification of  prequantizations}\label{Appendix}


\renewcommand{\theequation}{\thesection.\arabic{equation}}
\renewcommand
\appendix{\appendix}{\setcounter{equation}{0}}

\vskip-7mm
\kikezd{Souriau's classification \cite{JMS67,SSD}}

 Denote by 
$\big(\widetilde{E}, \pi_1,q\big)$ where $q$ is  the projection $\widetilde{E}\to E$ the universal covering of $E$ and define $\tilde{\sigma}  = q^*\sigma$. $\pi_1$, the first homotopy group of $E$, acts  on $\widetilde{E}$ by symplectomorphisms.

Let us choose a reference prequantisation $\big(Y_0, \omega_0, \pi_0\big)$ of $(E, \sigma)$. As 
$\big(\widetilde{E},\tilde{\sigma}\big)$ 
is simply connected, it has a unique prequantisation 
$\big(\widetilde{Y}_0, \widetilde{\omega}_0,\widetilde{\pi}_1\big)$ \cite{Kostant,SSD,JMS67},  which can be obtained from 
$\big(Y_0, \omega_0, \pi_0\big)$ by pull-back as
\beq
\big(\widetilde{Y}_0, \widetilde{\omega}_0,\widetilde{\pi}_1\big) = q^*\big(Y_0, \omega_0, \pi_0\big)\,.
\label{tildepreq}
\eeq

If $\chi=\pi_1 \to U(1)$ is a character of the first homotopy group, then $\pi_1$ admits an \emph{isomorphic lift} to $\big(\widetilde{Y}_0, \widetilde{\omega}_0,\widetilde{\pi}_1\big)$ of the form
\beq
\widehat{g}^{\chi} = \big(\tilde{x},\tilde{\xi}\big)  = \Big(g(\tilde{x}), \underline{\chi(g)}_{Y_0}(\xi)\Big)\,,
\label{charact}
\eeq
where $g\in\pi_1$ and $\underline{Z}_{Y_0}$ denotes, as below  \eqref{equivwf}, the action of $Z\in U(1)$ on $Y_0$ .
Now Souriau has shown that
\beq
\big(Y_{\chi}, \omega_{\chi}, \pi_{\chi}\big) =
\big(\widetilde{Y}, \widetilde{\omega},\widetilde{\pi}\big)/\hat{\pi}_1^{\chi}\,
\label{chiquant}
\eeq
is a prequantisation of $(E,\sigma)$, and that all  prequantisations are obtained in this way which proves Proposition \ref{prequantchar}:

\textit{
The inequivalent prequantisations are in 1-1 correspondence with the characters of the first homotopy group.}
 
\kikezd{A general scheme \cite{ackKollar}}

Now we outline the proof of the general classification  Theorem \ref{classifthm}.

\begin{prop}
If the homotopy group is finite, $|\pi_1| < \infty$, then $H^1(E,\IR)= 0$, \ie, every closed $1$-form is exact.
\label{H1=0}
\end{prop}

\noindent{Proof} : Let $\alpha$ be a closed $1$-form on $E$, put $\tilde{\alpha} = q^*\alpha$.  $\tilde{\alpha} = d\tilde{f}$ for $\widetilde{E}$ is simply connected. Define
$$
\tilde{h} = \frac{1}{|\pi_1|} \sum_{g\in\p_1} g^*f\,.
$$
$\tilde{h}$ is invariant under $g\in\pi_1$ and projects thus to $h: E\to \IR$. But $\tilde{\alpha}=d\tilde{h} = q^*\alpha$, and thus $\alpha = dh$.
The general situation can be treated by algebraic topology \cite{McLane}. Consider the exact sequence of groups 
\beq
0 \longrightarrow \IZ \xrightarrow{~2\pi i ~} 
U(1) \longrightarrow 0
\label{groupseq}
\eeq
giving rise to the long exact sequence
\beq\hskip-8mm
\longrightarrow H^1(E,\IZ) \xrightarrow{~\underset{%
\smile }{1}~} 
\underset{closed/exact}{H^{1}(E,\IR)}
\xrightarrow {~\underset{%
\smile }{2}~}
\underset{characters}{H^{1}(E,U(1))}
\xrightarrow {~\underset{%
\smile }{3}~}
\underset{Chern~class}{H^{2}(E,\IZ)}
\xrightarrow {~\underset{%
\smile }{4}~}
\underset{
\begin{array}{c}
curv.~class \\
\big[\sigma/2\pi\hbar\big]
\end{array}%
}
{H^{2}(E,\IR)}
\longrightarrow
\label{longsequence}
\eeq
Then we make the following observations:

\begin{enumerate}
\item $\sigma$ defines, by \eqref{prequantcond} an integer-valued element of
$ H^2_{dR}(E,\IR)$  which, by de Rham's theorem is just 
$H^2(E,\IR)$.

\item
The bundle topology is characterised by its Chern class which is in $H^2(E,\IZ)$. Thus we have as many bundles as elements in the kernel of $\underbar{4}$.

\item
As $\UU(1)$ is commutative, a character of $\pi_1$ depends only on $\pi_1/[\pi_1,\pi_1]$, which is known to be $H_1(E,\IZ)$. On the other hand, the Theorem of Universal Coefficients \cite{McLane} p.76 yields that
\beq
H^1(E,U(1)) \simeq 
{\rm Hom}\Big(H_1(E,\IZ\big), U(1)\Big) \,.
\label{Hom}
\eeq
Thus $H^1(E,U(1))$ is the \emph{set of all characters} 
which classify the different prequantisations as found by Souriau \cite{SSD}.

\item
Under quite general conditions, we have
\beq
\barraynb{lllll}
H^i(E,\IZ) &\simeq & \IZ^{b_i} &\oplus &\text{Tors} H^i
\\[4pt]
H_i(E,\IZ) &\simeq & \IZ^{b_i} &\oplus &\text{Tors} H_i
\earraynb
\label{HiHi}
\eeq
where $b_i$ is the Betti number \cite{Betti}. {Tors} $H^i$ and {Tors} $H_i$  are groups whose elements are all of finite order; 
\item
The kernel of the map $H^i(E,\IZ) \longrightarrow H^i(E,\IR)$ is just {Tors} $H^i$; the image of $\IZ^{b_i}$ is a basis in $H^i(E,\IR)$;
\item
Again by the Theorem of Universal Coefficients,
\beq
\text{Tors}\, H^2(E,\IZ) \simeq \text{Tors}\, H_1(E,\IZ)
\simeq \text{Tors}\, \pi_1/[\pi_1,\pi_1]\,.
\label{(20)}
\eeq
Thus by 2.) 5.), 6.),

\begin{prop}
The topologically distinct prequantum bundles are labelled by the elements of the Torsion subgroup in \eqref{(20)}.
\label{bundleclass}
\end{prop}

\item
According to 5.), the image of $H^1(E,\IZ)$ in $H^1(E,\IR)$
under 
1.) is composed of integer multiples of a basis. Thus 
$H^1(E,\IR)/({\rm Im}) \big(H^1(E,\IR)\big) \simeq
(S^1)^{b_1}$ and we get the exact sequence
\beq
0 \longrightarrow {S^1} \xrightarrow{b_{1}} H^1(E,U(1))
 \longrightarrow  \text{Tors} H_1(E,\IZ) \longrightarrow 0\,.
\label{exactsequence}
\eeq
Now by de Rham's theorem, to any element of $H^1(E,\IR)$
we can associate a closed $1$-form $\alpha/(2\pi\hbar)$ such that its value on $g\in H_1(E,\IR)$ is
\beq
\frac{1}{2\pi\hbar} \oint_{\gamma}\alpha\,,
\label{(22)}
\eeq
where the homology class of $\gamma$ is  $g$.

Next by \eqref{exactsequence}, the image of $(S^1)^{b_1}$ in $H^1(E,U(1))$ is composed of characters of the form
\beq
\chi(g) = \exp \left[\frac{i}{\hbar}\oint_{\gamma}\alpha\right]\,.
\label{(23)}
\eeq
As $(S^1)^{b_1}$ is connected and Tors\,$H^2(E,\IZ)$ is finite, we have~:

\begin{prop}
The characters of the form \eqref{(23)} make up the \emph{connected component} which contains the character $\chi\equiv 1$.
\label{concompchar}
\end{prop}
\item
Let us choose a basis of $H_{dR}^1(E,\IR)$ composed of the $1$-forms $\alpha_1,\dots, \alpha_{b_1}$, and pick  a character $\chi_k \in H^1(E,U(1))$ in each element of 
Tors\,$H^2$ = Tors\, $H_1$.

\eenu


\begin{thebibliography}{99}

\bibitem{HPAix79}
P.~A.~Horvathy,
``Classical Action, the {Wu-Yang} Phase Factor and Prequantization,''
Lect. Notes Math. \textbf{836} (1980), 67-90
doi:10.1007/BFb0089727

\bibitem{Kostant}
B.~Kostant,
``On certain unitary representations which arise from a quantization theory,''
Conf. Proc. C \textbf{690722} (1969), 237-253
doi:10.1007/3-540-05310-7\_28

\bibitem{SSD}
J.-M. Souriau,
\textsl{Structure des syst\`emes dynamiques}, Dunod (1970, \copyright\,1969);
\textsl{Structure of Dynamical Systems. A Symplectic View of Physics},
translated by C.H.~Cushman-de Vries (R.H.~Cushman and G.M.~Tuynman, Translation Editors), Birkh\"auser, 1997.
The ``monopole without strings'' discussions  \cite{WuYangLag,BalaMarmo} are discussed also in the Prequantization Chap.~V. written around 1975 for the planned  but never completed revised edition of Souriau's book.

\bibitem{SimmsWoodhouse}
D.~J.~Simms and N.~M.~J.~Woodhouse,
``Lectures on Geometric Quantization,''
Lect. Notes Phys. \textbf{53} (1976), 1-166
doi:10.1007/3-540-07860-6

\bibitem{DowkerAustin}
J.S. Dowker,
{\sl ``Selected topics in topology and quantum field theory,''}, Austin Lectures, Jan - May 1979.

\bibitem{ackKollar}
The results presented here were obtained in collaboration with J. Koll\'ar.

\bibitem{Asorey}
M.~Asorey,
``Some Remarks on the Classical Vacuum Structure of Gauge Field Theories,''
J. Math. Phys. \textbf{22} (1981), 179
[erratum: J. Math. Phys. \textbf{25} (1984), 187]
doi:10.1063/1.524732

\bibitem{FriedmanSorkin}
J.~L.~Friedman and R.~D.~Sorkin,
``DYON SPIN AND STATISTICS: A FIBER BUNDLE THEORY OF INTERACTING MAGNETIC AND ELECTRIC CHARGES,''
Phys. Rev. D \textbf{20} (1979), 2511-2525
doi:10.1103/PhysRevD.20.2511
J.~L.~Friedman and R.~D.~Sorkin,
``A SPIN STATISTICS THEOREM FOR COMPOSITES CONTAINING BOTH ELECTRIC AND MAGNETIC CHARGES,''
Commun. Math. Phys. \textbf{73} (1980), 161-196
doi:10.1007/BF01198122

\bibitem{SimmsAix79}
D.~J.~Simms,
``GEOMETRIC ASPECTS OF THE FEYNMAN INTEGRAL,''
Lect. Notes Math. \textbf{836} (1980), 167-170
See also
D.~J.~Simms,
``Geometric Quantization and the Feynman Integral.''
Contribution to: {\sl Mathematical Problems in Feynman Path Integral, 220-223}.

\bibitem{Laidlaw}
M.~G.~G.~Laidlaw and C.~M.~DeWitt,
``Feynman functional integrals for systems of indistinguishable particles,''
Phys. Rev. D \textbf{3} (1971), 1375-1378
doi:10.1103/PhysRevD.3.1375

\bibitem{McLane}
S. McLane, {\sl Homology}, Springer (1967)


\bibitem{JMS67}
J-M Souriau
``Quantification g\'eom\'etrique. Applications.''
Ann. Inst. H. Poincar\'e Sect. A (N.S.)  6  311-341  (1967)

\bibitem{Schulman}
L.~Schulman,
``A Path integral for spin,''
Phys. Rev. \textbf{176} (1968), 1558 -1569
doi:10.1103/PhysRev.176.1558.
L. S. Schulman,
``Approximate topologies''
J. Math. Phys. \textbf{12}, 304 (1971)
https://doi.org/10.1063/1.1665592

\bibitem{Dowker72}
J.~S.~Dowker,
``Quantum mechanics and field theory on multiply connected and on homogeneous spaces,''
J. Phys. A \textbf{5} (1972), 936-943
doi:10.1088/0305-4470/5/7/004

\bibitem{SchulmanPathInt}
L. S. Schulman,
{\sl Techniques and applications of path integration}, J. Wiley, N.Y. (1981)

\bibitem{HPAPLA}
P.~A.~Horvathy,
``Quantization in Multiply Connected Spaces,''
Phys. Lett. A \textbf{76} (1980), 11-14
doi:10.1016/0375-9601(80)90133-4

\bibitem{HPClausthal}
P.~A.~Horvathy,
``Prequantisation From Path Integral Viewpoint,''
Lect. Notes Math. \textbf{905} (1982), 197-206
CPT-80-P-1230.

\bibitem{HoMoSu}
P.~A.~Horvathy, G.~Morandi and E.~C.~G.~Sudarshan,
``Inequivalent quantizations in multiply connected spaces,''
Nuovo Cim. D \textbf{11} (1989), 201-228
doi:10.1007/BF02450240

\bibitem{Dirac33}
P.~A.~M.~Dirac,
``The Lagrangian in quantum mechanics,''
Phys. Z. Sowjetunion \textbf{3} (1933), 64-72

\bibitem{Feynman42}
R.~P.~Feynman,
``The principle of least action in quantum mechanics,''
doi:10.1142/9789812567635\_0001

\bibitem{FeynmanHibbs}
R. P. Feynman and A. R. Hibbs,
{\sl Quantum Mechanics and Path Integrals}. McGraw-Hill, New York N.Y., (1965)

\bibitem{Torino81}
P.~A.~Horvathy,
``AN ACTION PRINCIPLE FOR ISOSPIN,''
BI-TP-82-19.
Proceedings of the IUTAM-ISIMM Symposium on Modern Developments in Analytical Mechanics, Torino, 982, Atti Acad. Sci. Torino, Suppl. \textbf{117}, 163-169
Accademia delle scienze (1983).

\bibitem{WuYangLag}
T.~T.~Wu and C.~N.~Yang,
``Dirac's Monopole Without Strings: Classical Lagrangian Theory,''
Phys. Rev. D \textbf{14} (1976), 437-445
doi:10.1103/PhysRevD.14.437

\bibitem{BalaMarmo}
A.~P.~Balachandran, G.~Marmo and A.~Stern,
``Magnetic Monopoles With No Strings,''
Nucl. Phys. B \textbf{162} (1980), 385-396
doi:10.1016/0550-3213(80)90346-6,
A.~P.~Balachandran, G.~Marmo, B.~S.~Skagerstam and A.~Stern,
``Supersymmetric Point Particles and Monopoles With No Strings,''
Nucl. Phys. B \textbf{164} (1980), 427-444
[erratum: Nucl. Phys. B \textbf{169} (1980), 547]
doi:10.1016/0550-3213(80)90520-9

\bibitem{Betti}
https://en.wikipedia.org/wiki/Betti${}_{}$ number

\bibitem{AsoreyBoya}
M.~Asorey and L.~J.~Boya,
``ELECTROMAGNETISM WITHOUT MONOPOLES IS POSSIBLE IN NONTRIVIAL U(1) FIBER BUNDLES,''
J. Math. Phys. \textbf{20} (1979), 2327
doi:10.1063/1.524013

\bibitem{Leinaas77}
J.~M.~Leinaas and J.~Myrheim,
``On the theory of identical particles,''
Nuovo Cim. B \textbf{37} (1977), 1-23
doi:10.1007/BF02727953

\bibitem{Wilczek82}
F.~Wilczek,
``Quantum Mechanics of Fractional Spin Particles,''
Phys. Rev. Lett. \textbf{49} (1982), 957-959
doi:10.1103/PhysRevLett.49.957
Wilczek, Frank (2021). {\sl Fundamentals : Ten Keys to Reality}. New York, New York: Penguin Press. pp. 89-90. ISBN 9780735223790. LCCN 2020020086

\bibitem{HP-Kollar}
P.~A.~Horvathy and J.~Kollar,
``The Nonabelian {Aharonov-Bohm} Effect in Geometric Quantization,''
Class. Quant. Grav. \textbf{1} (1984), L61
doi:10.1088/0264-9381/1/6/002.

\end{thebibliography}
\end{document}